\documentclass[12pt, draftclsnofoot, onecolumn]{IEEEtran}
\usepackage{amsmath,amsthm,amsfonts,amscd,amssymb,graphics,dsfont}
\usepackage[mathscr]{eucal}
\usepackage{graphics,graphicx,multicol}
\usepackage{epsfig}
\usepackage{epstopdf}
\usepackage{subfigure}
\usepackage{stfloats}
\usepackage{latexsym}
\usepackage{amsfonts}
\usepackage{cite}
\usepackage{algorithm,algorithmic}
\usepackage{color,xcolor}
\usepackage{multirow}

\begin{document}
\title{Machine Learning-Aided Cooperative Localization under Dense Urban Environment}
\author{Hoon Lee,%~\IEEEmembership{Member,~IEEE}, 
~Hong Ki Kim, Seung Hyun Oh, and Sang Hyun Lee%,~\IEEEmembership{Member,~IEEE}
%, and Tony Q. S. Quek,~\IEEEmembership{Fellow,~IEEE}
\thanks{
H. Lee is with the Department of Electrical Engineering and the Artificial Intelligence Graduate School, Ulsan National Institute of Science and Technology (UNIST), Ulsan, 44919, Korea. % (e-mail: hlee@pknu.ac.kr).

H. K. Kim, S. H. Oh, and S. H. Lee are with the School of Electrical Engineering, Korea University, Seoul 02841, Korea (e-mail: sanghyunlee@korea.ac.kr).

%T. Q. S. Quek is with the Information Systems Technology and Design Pillar, Singapore University of Technology and Design, Singapore 487372 (e-mail: tonyquek@sutd.edu.sg).
}}

\maketitle
%%%%%%%%%%%%%%%%%%%%%%%%%%%%%%%%%%%%%%%%%%%%%%%%%%%%%%%%%%%%%%%%%%%%%%%%%%%%%%%%%%%%%%%%%%%%%%%%%%%%%%%%Abstract
\begin{abstract}
Future wireless network technology provides automobiles with the connectivity feature to consolidate the concept of vehicular networks that collaborate on conducting cooperative driving tasks. The full potential of connected vehicles, which promises road safety and quality driving experience, can be leveraged if machine learning models guarantee the robustness in performing core functions including localization and controls. Location awareness, in particular, lends itself to the deployment of location-specific services and the improvement of the operation performance. The localization entails direct communication to the network infrastructure, and the resulting centralized positioning solutions readily become intractable as the network scales up. As an alternative to the centralized solutions, this article addresses decentralized principle of vehicular localization reinforced by machine learning techniques in dense urban environments with frequent inaccessibility to reliable measurement. As such, the collaboration of multiple vehicles enhances the positioning performance of machine learning approaches. A virtual testbed is developed to validate this machine learning model for real-map vehicular networks. Numerical results demonstrate universal feasibility of cooperative localization, in particular, for dense urban area configurations.
\end{abstract}

\section{Introduction}

Rapid advances in automobile industry and information technology transform cars from traditional means of transportation to information-oriented commuting machines on the go \cite{Hussain:19}. Outfitting automobiles with wireless capabilities yields proactive and cooperative connection solutions that enable on-board network access, energy-cost reduction, and location-dependent services \cite{Lu:14}. Efficient coupling of such features also provides foundations for the realization of autonomous driving with promise of improving road safety and transporting mobility-impaired populations. 
Vehicles leverage vehicle-to-everything (V2X) communication to share information in real-time, and their basic operations can be distributed among neighboring vehicles \cite{Balkus:22}. Public interest in such services is booming, and their market potential is projected to reach nearly \$800 billion by 2050~\cite{Yurtsever:20}. 

Location awareness, realized by identifying physical ego-positions of vehicles, enables to launch service of location-based applications and places fundamental requirements for global navigation \cite{Safavi:18}. 
The integration of mobile objects improves connectivity and coverage of the network to provide essential information since additional measurements can be collected. 
For ultra-reliable positioning, intensive research has been studied on localization via built-in V2X infrastructure \cite{SWK:21,Henk:17}.
The nature of localization solutions depends on what types of measurements are acquired and how measurements are processed whether by central computing units or in decentralized fashion using in-vehicle computing. 

Global navigation satellite systems (GNSSs) have been widely adopted for vehicular localization. However, GNSS measurements are often inaccessible in harsh environments, in particular, urban areas of densely located signal-blocking obstacles. Although dead reckoning in time intervals with absolute GNSS readings envisions to calculate current positions of vehicles, its uncertainty on blockage limits the localization accuracy. Thus, efficient and low-cost localization does not necessarily rely on direct access to network infrastructure nor exploit costly LIDAR sensing equipment. V2X-based localization reuses wireless propagation measurements obtained at vehicles \cite{DJin:20}. 

Localization over vehicular networks is essentially a time-varying mission since vehicular locations change during every run. Mobility of vehicular objects makes positioning more challenging in that their locations and available neighborhood reached by peer-to-peer connectivity keep changing. Therefore, the design of rapid and accurate localization is critical without compromising quality for embedded applications requiring real-time position information. 

Machine learning (ML) algorithms become backbones of contemporary V2X networks in handling computation-intensive services resulting from time-varying road environments.
Connected vehicles rely on neural network models to address major driving missions: perception, maneuver, 
and route planning. Accurate algorithms involve the collection of informative data, and high-resolution localization is also possible via central collection of local measurements from vehicles. Such positioning computations, however, may not be feasible due to lack of guaranteed central coordination and high-reliability anchors. 
To cope with challenging situations, decentralized mechanisms are essential for individual vehicles to locate themselves without centralized processing. With the aid of V2X networks, sensor measurement is kept on vehicles and only locally shared upon request by neighbors. This potentially saves communication costs for continuous streams of sensor-monitored information, thereby sparing vehicular service performance. The resulting localization task is posed as a decentralized optimization. To address network management challenges, an ML framework of vehicular collaboration is presented, in particular, tailored to ML-aided collaborative localization (MLCL) tasks.

This article discusses localization challenges for vehicular networks and presents a cooperation principle for identifying locations and mobilities of individual vehicles. Subsequently, MLCL formalism is developed as decentralized learning tasks.
Mobile behaviors and measurements are monitored via virtual testbed developed over urban-map environments. ML-based solutions are tested through comparison with existing techniques. Finally, the article is concluded with challenges and future perspectives of cooperative ML solutions.

\section{Localization in vehicular networks}

State-of-the-art sensing equipment, such as GNSS sensors, odometers, and accelerometers, enables individual vehicles for independent localization with their measured information about travel distance, acceleration, and orientation. The proximity of nested vehicular mobility entails correlated internal measurements among vehicles. Such internal measurements are prone to sensing errors, which become severe in dense urban environments. Furthermore, GNSS measurements become even unavailable when GNSS signals are occluded by overpasses or tunnels. The resulting accuracy of independent localization often fails to meet requirements of high-level vehicular applications, which triggers the development of cooperative mechanisms for reliable localization. This section introduces principles for cooperative localization and discusses key enablers of decentralized techniques.

\subsection{Vehicular network models}

\begin{figure*}
\centering
\includegraphics[width=\linewidth]{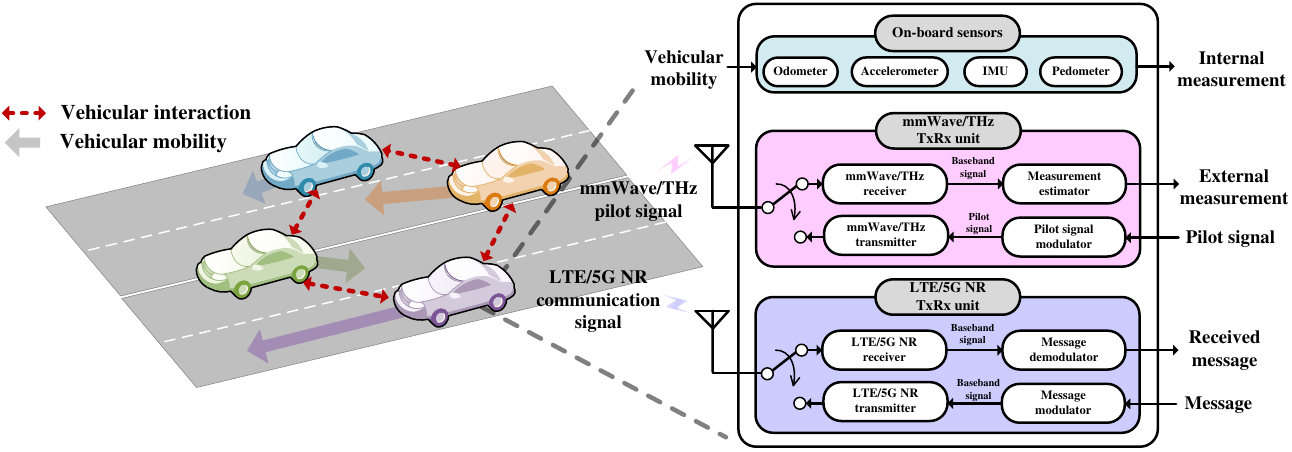}
\caption{Cooperative localization over vehicle group.}
\label{fig:fig1}
\end{figure*}

Fig. \ref{fig:fig1} illustrates cooperative localization tasks on roadway configurations. Multiple vehicles desire to identify individual positions without central network intervention. In addition to internal measurements obtained using on-board equipment, vehicles observe additional measurements from neighboring vehicles. 
Such external measurements contain distances and relative orientations of neighboring vehicles. They are acquired by processing relevant statistics, e.g., time-of-arrival, angle-of-arrival, and received signal strength \cite{DJin:20}. Individual vehicles can obtain these statistics passively by sensing positioning reference signals (PRSs) transmitted from neighbors over optical or high-frequency radio transmission, e.g., mmWave/THz bands \cite{Henk:17,SWK:21}.
Passively observed measurements, therefore, do not convey informative and proactive knowledge about internal measurements, i.e., mobility and position estimates of neighbors. For active vehicular interactions through V2X connections, coordination protocols exchange relevant statistics encoded as communication messages among vehicles over reliable LTE/5G links. Connected vehicles are configured via standard initialization protocols, such as 3GPP LTE small-size messaging protocols to facilitate proactive inter-vehicle interaction. Message generation primarily depends on measurements, driving environments, and channel propagation conditions. The resulting localization task conceives joint optimization of interaction strategies and position computations.

\subsection{Fundamental challenges}
The design of cooperative localization poses the following fundamental challenges.

\subsubsection{Model dependence}
Existing cooperative algorithms \cite{DJin:20,HKim:22,Kia:16} resort to perfect knowledge of vehicle statistics, e.g., probability distributions about positions and measurement qualities. In practice, time-varying vehicular network dynamics limit access to their statistical information, and model-dependent algorithms are hardly permitted to extensions to diverse network configurations. Model-free localization with data-driven ML strategies is essentially approached.

\subsubsection{Adaptive vehicular interactions}  
The channel sparsity of mmWave/THz bands is beneficial for evaluating LoS-dependent external measurements. The availability of LoS-based vehicular connections is, however, restricted by vehicular mobility and radio propagation characteristics. Insufficient connections for vehicular pairs suffer from unreliable external measurements and communicating messaging over wireless transmission.
Thus, universal localization rules necessarily adapt to vehicular link connectivity.

\subsubsection{Heterogeneous vehicular connectivity}
For reliable message transmission, vehicular links are established via LTE/5G networks. Communication messages undergo radio propagation distinct from external measurements. This brings forth heterogeneous vehicular interactions with measurement pilots and communication messages.
An aggregation strategy combining unbalanced information from heterogeneous vehicular interconnections is essential.

\subsubsection{Time-varying mobility}
Individual vehicles are navigated through route-planning algorithms. However, actual driving performance depends on time-varying traffic and roadway conditions, e.g., traffic signals and congestion, and the resulting trajectories are strongly correlated with nearby vehicles. Non-stationary traffic flows exhibit non-Markovian dynamics. Such vehicular behaviors solicit self-organizing localization mechanisms that universally operate over arbitrary driving configurations.

\section{Cooperative localization strategy}

\begin{figure*}
\centering
\includegraphics[width=\linewidth]{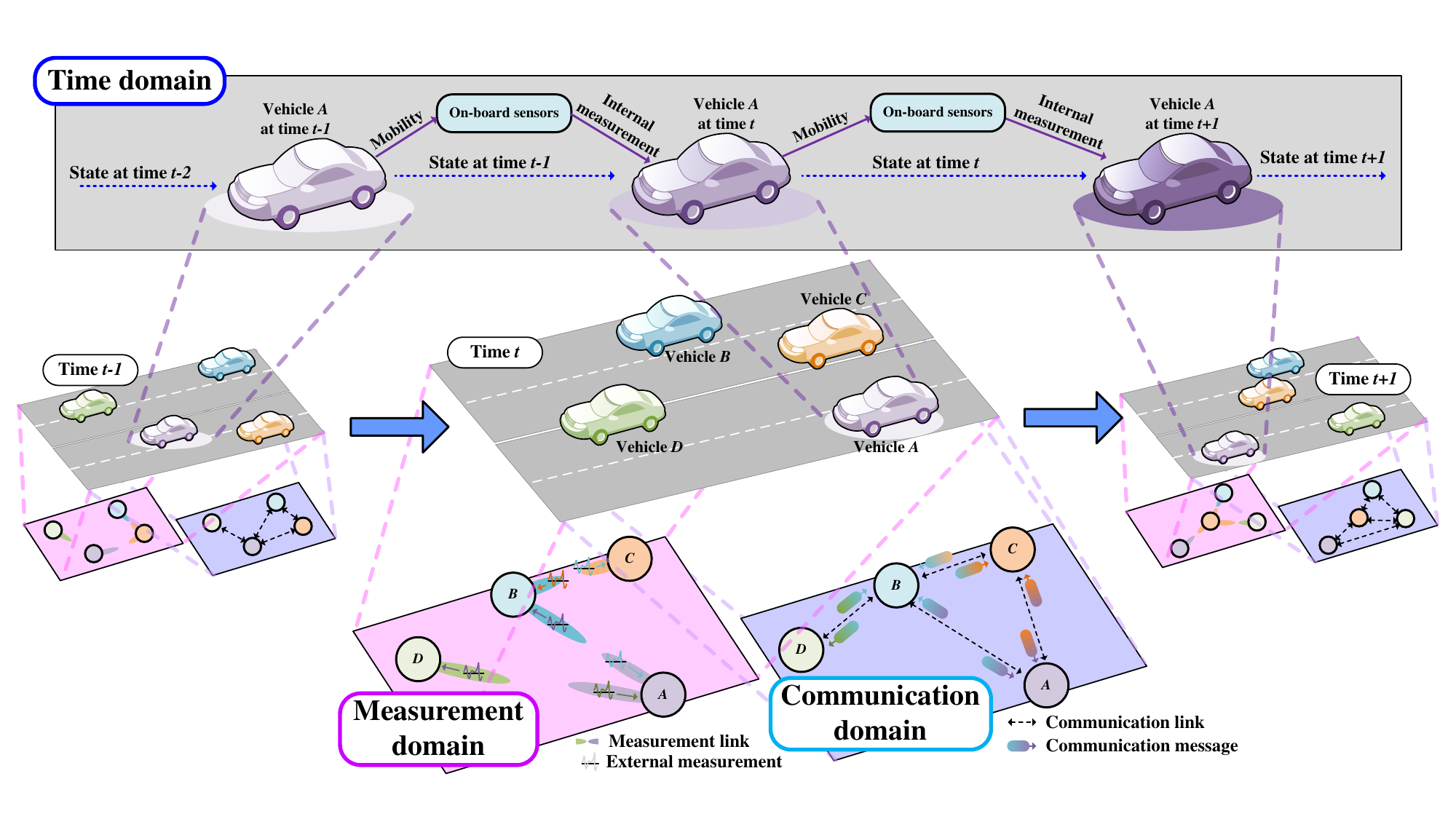}
\caption{Vehicle interactions over three domains of the MLCL framework.}
\label{fig:fig2}
\end{figure*}

To address the aforementioned challenges, cooperative localization strategies are organized with nested vehicle-to-vehicle cooperation in several interaction domains. This is a key enabler in developing viable ML models for cooperative vehicular localization.

\subsection{Vehicular interaction domains}
Figure \ref{fig:fig2} illustrates three-level vehicular interactions slicing vehicular networks: measurement, communication, and time domains.

\subsubsection{Measurement domain} 
This domain is engaged in inter-vehicle coordination to collect external measurements, including distances and relative orientations of neighboring vehicles. Vehicles connected in measurement domain receive PRSs transmitted over mmWave/THz bands to obtain external measurements.
This is characterized via a graph where vehicles are interpreted as vertices and their edge connections represent the reciprocal availability of their external measurements.
    
\subsubsection{Communication domain} 
Communication domain establishes proactive vehicular interactions via the exchange of communication messages over LTE/5G links.
The corresponding graphical model is represented with vehicle-to-vehicle communication connections.  
Heterogeneous wireless propagation environments entail inter-vehicle interactions independent of measurement-domain connections. 
Its primary design goal is to develop model-free ML models for encoding communication messages to encapsulate collected external measurements among vehicles.

\subsubsection{Time domain}
Vehicular interactions are also subject to time-varying road traffic variations and network topology dynamics. Time domain is responsible for capturing temporal correlations of vehicular dynamics with internal measurement obtained from on-board sensors. 
This domain propagates current intra-vehicle state information about internal/external measurement and communication messages to the next localization time instant so that past positioning history is exploited for future localization estimates. ML approaches are ideal enhancements for handling such historical knowledge by means of recurrent neural architecture.

\begin{table*}[]
\caption{Comparison of localization techniques ($\bigcirc$: positive, $\bigtriangleup$: neutral, $\times$: negative)}
\label{tab:tab1}
\centering
\resizebox{\textwidth}{!}{
\begin{tabular}{|c||c|c|c|c|c|c|}
\hline
Category                       & Techniques                                                   & \begin{tabular}[c]{@{}c@{}}Decentralized\\ implementation\end{tabular} & \begin{tabular}[c]{@{}c@{}}Model-independent\\ processing \end{tabular} & \begin{tabular}[c]{@{}c@{}}Adaptive\\ interaction\end{tabular} & \begin{tabular}[c]{@{}c@{}}Heterogeneous\\ connectivity\end{tabular} & \begin{tabular}[c]{@{}c@{}}Time-varying\\ mobility\end{tabular} \\ \hline
\multirow{4}{*}{Model-based}   & SPAWN \cite{DJin:20}                                                       & $\bigcirc$                                                                       & $\times$                                                                      & $\times$                                                                  & $\times$                                                                    & $\bigcirc$                                                               \\ \cline{2-7} 
                               & \begin{tabular}[c]{@{}c@{}}Maximum\\ likelihood \cite{HKim:22}\end{tabular} & $\bigcirc$                                                                        & $\times$                                                                     & $\times$                                                                  & $\times$                                                                    & $\bigcirc$                                                               \\ \cline{2-7} 
                               & EKF \cite{Kia:16}                                                         & $\bigcirc$                                                                     &  $\times$                                                                       & $\times$                                                                  & $\times$                                                                    & $\bigcirc$                                                               \\ \hline
\multirow{2}{*}{Surrogate ML}  & \begin{tabular}[c]{@{}c@{}}Learnable\\ SPAWN\cite{JEom:19} \end{tabular}    & $\bigcirc$                                                             & $\triangle$                                                                      & $\times$                                                                  & $\times$                                                                    & $\times$                                                               \\ \cline{2-7} 
                               & KalmanNet \cite{GRevach:22}                                                   & $\times$                                                             & $\triangle$                                                                      & $\times$                                                                  & $\times$                                                                    & $\bigcirc$                                                               \\ \hline
\multirow{3}{*}{Model-free ML} & GCN \cite{Lin,WYan:21}                                                         & $\times$                                                                     & $\bigcirc$                                                                        & $\bigcirc$                                                                  & $\times$                                                                    & $\times$                                                               \\ \cline{2-7} 
                               & \begin{tabular}[c]{@{}c@{}}MLCL\\ (Proposed)\end{tabular}    & $\bigcirc$                                                                       & $\bigcirc$                                                                      & $\bigcirc$                                                                  & $\bigcirc$                                                                    & $\bigcirc$                                                               \\ \hline
\end{tabular}
}
\end{table*}

\subsection{State-of-the-arts of localization techniques}

Advanced localization frameworks can be consolidated through the orchestration of measurement, communication, and time domains. Such domain-integrated design principles address fundamental challenges of real-world vehicular networks, i.e., model dependency, adaptive interactions, heterogeneous connectivity, and time-varying mobility, as discussed in the previous section. Therefore, cooperative localization techniques can be investigated in terms of managing three domains, as summarized in Table \ref{tab:tab1}.

Several localization solutions, including sum-product algorithm over wireless networks (SPAWN) \cite{DJin:20}, decentralized maximum likelihood \cite{HKim:22}, and decentralized extended Kalman filter (EKF) \cite{Kia:16}, guarantee certain levels of convergence and performance under well-defined models of sensing quality in measurement domain and vehicular mobility in time domain. As such, they are referred to as \textit{model-based} approaches. These frameworks incorporate homogeneous propagation over measurement and communication domains, provided that measurement pilots and communication messages are shared over identical wireless media. 
Therefore, all vehicular pairs are necessarily interconnected in both domains, thereby failing to capture adaptive and heterogeneous inter-vehicle interactions.

ML techniques can circumvent such difficulties of model-based frameworks \cite{JEom:19,GRevach:22}. Deep neural network (DNN) models construct \textit{surrogate models} substituting model-dependent calculations in communication and time domains. Surrogate DNNs are trained with data samples collected from model-based approaches. Trained DNNs are straightforwardly tweaked into real-time inference calculations for model-based localization. Learnable SPAWN \cite{JEom:19} replaces the SPAWN communication domain with fully-connected neural networks (FNNs). 
FNN models are trained to determine vehicular positions with noisy external measurements, and external measurement processing is exempted from prior knowledge about noise statistics. FNN outputs are shared with neighbors over communication domains, and vehicular positions are estimated with average incoming messages. KalmanNet \cite{GRevach:22} employs recurrent neural networks (RNNs) that conduct time-domain Kalman gain calculations. Although its efficient Kalman gain calculation without measurement processing is evidenced in sequential estimation tasks, it relies on costly global collection of measurement inputs and intensive centralized computations. Also, KalmanNet requires reliable models for vehicular mobility. Since 
surrogate models address only a subset of interaction domains of model-based localization with DNN techniques, shortcomings of the original model-based localization, e.g. heterogeneous connectivity and time-varying inter-vehicle interactions, are also inherited.

{\em Model-free} ML designs merge individual steps of position computation processing in a single DNN arrangement \cite{WYan:21,Lin}, which relieves, by consequence, prior knowledge requirements about vehicular mobility and measurement. 
Under network environments with noisy external measurements of inter-vehicle distances, the corresponding measurement domain is envisioned in a graph-based context.
Structured arrangements of data collections introduce graph convolutional networks (GCNs). These new models conduct convolution filtering among interacting vehicles in measurement domain to obtain simultaneous position estimates.

Despite existing model-free ML techniques, their DNN architecture approached through single-domain features still lacks the feasibility for real-world vehicular deployment. Conventional presumption about homogeneous measurement and communication domains comes from the configuration where external measurements and communication messages undergo an identical propagation environment. Such design principles, however, may fail to capture heterogeneous features from different coordination domains. In addition, missing time-domain consideration hinders the characterization of vehicle mobility and time-varying interactions in real-time localization \cite{WYan:21}, while global collection of local measurements is necessary for positioning computations \cite{Lin}.

\section{Proposed MLCL Framework}

\begin{figure*}
\centering
\includegraphics[width=\linewidth]{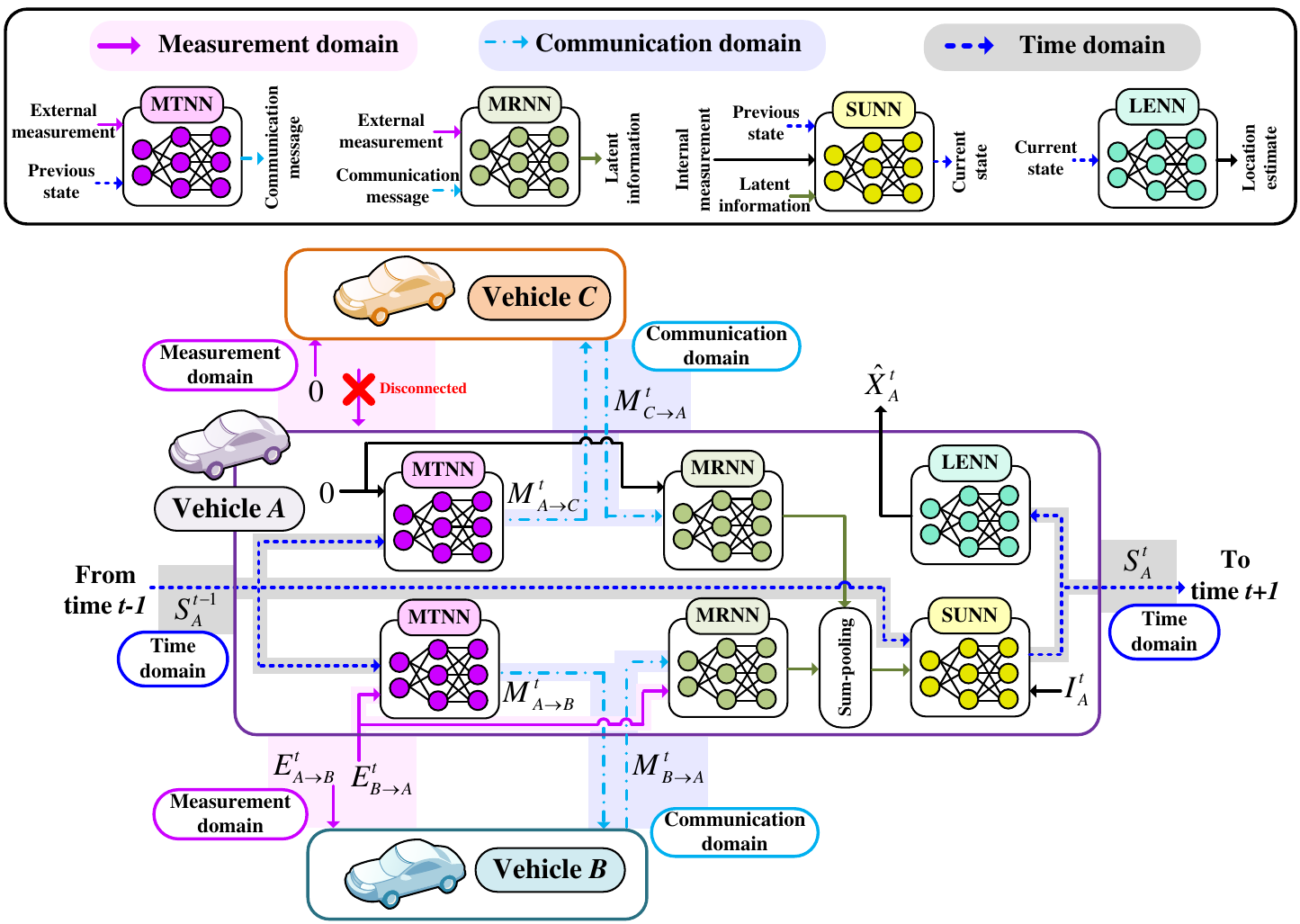}
\caption{Decentralized structure of MLCL models.}
\label{fig:fig3}
\end{figure*}

This section presents a model-free MLCL framework that integrates positioning calculations in measurement, communication, and time domains into an end-to-end learning architecture.

\subsection{Deep neural network structure}

Figure \ref{fig:fig3} shows the MLCL framework incorporating four core DNN units: message-transmit neural network (MTNN), message-receive neural network (MRNN), state-update neural network (SUNN), and location-estimate neural network (LENN), cooperatively responsible for measurement, communication, and time domain processing. These core DNN units have the following operations in cooperative localization as:
\begin{itemize}
    \item \textit{Message-transmit neural network}: Upon collection of external measurements in measurement domain, each vehicle generates communication messages to support inter-vehicle interactions in communication domain. Such a domain transformation is processed by MTNN that calculates messages with external measurement inputs.
    \item \textit{Message-receive neural network}: MRNN collects communication messages transferred from neighboring vehicles in communication domain and its own external measurements. This unit produces the output by extracting combined knowledge from measurement and communication domains.
    \item \textit{State-update neural network}: SUNN is responsible for time domain processing of individual vehicles. This unit receives MRNN-generated outputs to convert them to states, which are leveraged as inputs of DNN units for the next position estimate. Therefore, SUNN holds the current-state information and propagates it to other units for future processing. 
    \item \textit{Location-estimate neural network}: Final estimate of an individual vehicle is evaluated by LENN using the current-state information transferred by SUNN. 
\end{itemize}

As shown in Fig. \ref{fig:fig3}, these four core DNNs between neighboring vehicles cooperatively interact in the following way.

\subsubsection{Message-transmit neural network}
MTNN encodes the information for use in communication domain. To this end, MTNN unit of vehicle $A$ combines its previous state $S_{A}^{t-1}$ and external measurement $E_{B\rightarrow A}^{t}$ about vehicle $B$ at time $t$. Provided that two vehicles are interconnected in communication domain, MTNN unit generates communication message $M_{A\rightarrow B}^{t}$ bound for vehicle $B$. 
An adaptive learning architecture to address insufficient vehicular interaction is realized by switching control for MTNN. If two vehicles $A$ and $B$ have no connection in measurement domain, a zero input is fed into external measurement input $E_{B\rightarrow A}^{t}$ to MTNN unit of vehicle $A$, which is supposed to be transferred from vehicle $B$. 
Furthermore, MTNN message $M_{A\rightarrow B}^{t}$ is transferred to vehicle $B$ along the link between vehicles $A$ and $B$ in communication domain. 

\subsubsection{Message-receive neural network}
MRNN unit of vehicle $A$ combines communication message $M_{B\rightarrow A}^{t}$ and external measurement $E_{B\rightarrow A}^{t}$ obtained from vehicle $B$. In case of no available communication link between two vehicles $A$ and $B$, communication message input $M_{B\rightarrow A}^{t}$ can also be replaced with a zero vector message, as in MTNN. Furthermore, independently generated connectivity induces heterogeneous inter-vehicle interactions over measurement and communication domains. Even though vehicles $A$ and $C$ are disconnected in measurement domain as shown in Fig. \ref{fig:fig3}, they can interact with each other in communication domain. The MRNN output, referred to as latent information, bears all knowledge about vehicle $B$ available at vehicle $A$. Thus, all MRNN outputs are organized by a sum-pooling layer to obtain input features for subsequent SUNN. Switching configurations in MTNN and MRNN units allow inter-vehicle connectivity variations in measurement and communication domains in forward- and backward-pass computations of the proposed ML model, so that DNN units are optimized over time-varying vehicular network topologies.
 
\subsubsection{State-update neural network}
SUNN unit encodes current state $S_{A}^{t}$ that conveys all knowledge collected at vehicle $A$ at time $t$. Thus, SUNN inputs include latent information obtained by MRNNs and internal measurement $I_{A}^{t}$ sensed from on-board sensors, e.g., noisy positioning estimates using GNSS signals. Also, previous state $S_{A}^{t-1}$ generated by SUNN at time $t-1$ is fed in as side information. This recurrent ML architecture allows SUNN to capture temporal correlations of individual vehicles, thereby facilitating time-domain processing. Then, current states become sufficient statistic inputs to LENN unit which determines final position estimates. Also, state information is held internally for position tracking at the next localization period.

\subsubsection{Location-estimate neural network}
LENN takes current state $S_{A}^{t}$ as its only input for position estimation. The position estimate of vehicle $A$, $\hat{X}_{A}^{t}$, is desirably close to its instantaneous position $X_{A}$. This goal is considered in the loss function for training the proposed framework. DNN models are trained to reduce errors in group position estimates using true positions.

\subsection{Neural parameter optimization}
A supervised training strategy can be used to optimize neural parameters of group DNN units. 
The training dataset is prepared with input feature collections (internal measurement and external measurements) and labels (ground-truth positions). 
For successive localization tasks, a proper construction of current state $S_{A}^{t}$ necessarily grasps latent features of intra-vehicle interactions. Therefore, the dataset is prepared with time-series samples during $T$ consecutive intervals. The mean-absolute-error (MAE) measure between labels and forward-pass results over $T$ time intervals are optimized for training loss function. No regularization and additional labels are applied for communication messages and states. This goal-oriented optimization drives DNN units to learn effective generation policies about communication messages and states for accurate localization.
Back-propagation through time (BTTT) algorithm propagates gradients in time-reverse directions for backward-pass computation. 
SUNN unit of time $t$ can observe gradients of one-step forward MAE at time $t+1$, which is transferred from LENN unit of time $t+1$. Subsequently, state calculation rule about $S_{A}^{t}$ is adjusted to reduce possible future positioning errors.

Group training computations proceed by minimizing the sum-loss function over all vehicular groups.
The training mechanism incorporates model-free optimization of inter-vehicle interactions in measurement and communication domains. In the forward-pass phase, MTNN and MRNN units exchange communication messages over time domain, and BTTT algorithm propagates gradient information among connected vehicles. Once MRNN unit receives gradient information about loss functions of adjacent vehicles, this information propagates SUNN and LENN units, reaching MTNN unit. Thus, inter-vehicle interaction policy is optimized jointly with respect to communication domain via MTNN unit and time domain via SUNN unit, respectively. Four core units are trained in a nested way across all domains so that communication messages generated by MTNN become informative for vehicular cooperation. 

Statistics of vehicular measurements are normally non-stationary, in particular, in urban areas. Individual vehicles equipped with distinct DNN units dedicated to them are prone to overfitting to local datasets, thereby suffering from poor adaptation to environmental variations. Thus, component DNN units trained in one vehicle cannot be transplanted directly to another vehicle that experiences different dynamics. To train universal adaptations, the core DNN suite is applied for a distinct vehicle on every training episode. 
This training strategy naturally scales up in terms of measurement, communication, and time domain operations. The trained vehicles can be deployed at brand-new network configurations with different vehicle populations and long-duration driving controls. 

Trained DNN units are installed at individual vehicles for real-time cooperative localization. The inference of the trained MLCL is processed by simple matrix multiplications. The trained MLCL yields the localization results within 8ms, which complies with vehicular network latency requirement of LTE/5G standards \cite{VNC18}.

\section{Numerical evaluation}
This section assesses localization performance of MLCL framework. The simulations are evaluated with real-map virtual platform over an urban environment.

\subsection{Urban map virtual testbed}

Virtual testbed has been developed on a dense urban environment with realistic inter-vehicle communication and localization configurations. A $1400$m $\times$ $1200$m campus area of Korea University, Seoul, is chosen for demonstrative purposes. Geographical objects, such as roadways and city infrastructures, are constructed with Open Street Map. Road objects include junctions and interconnected lanes. Average vehicular mobility has been collected on-site, and multiple vehicular instances have been bootstrapped with Simulation of Urban MObility (SUMO). Vehicular trajectories of 505 vehicles with random pairs of origin and destination are collected, where 354 and 151 vehicles are used in training and test datasets, respectively. A graphical user interface (GUI) has been developed using a graphical physics engine of Unity. Vehicle trajectories are obtained with the combination of shortest-path route planning between arbitrary departure and arrival points and random turns after reaching the destination. Mobility dataset has been created in multiple batches to prevent improbably dense traffic that causes indefinite congestion. 
Vehicles contained in the same batch interact and interfere with the mobility of each other to reflect realistic traffic patterns. 

\begin{figure*}
\centering
\includegraphics[width=\linewidth]{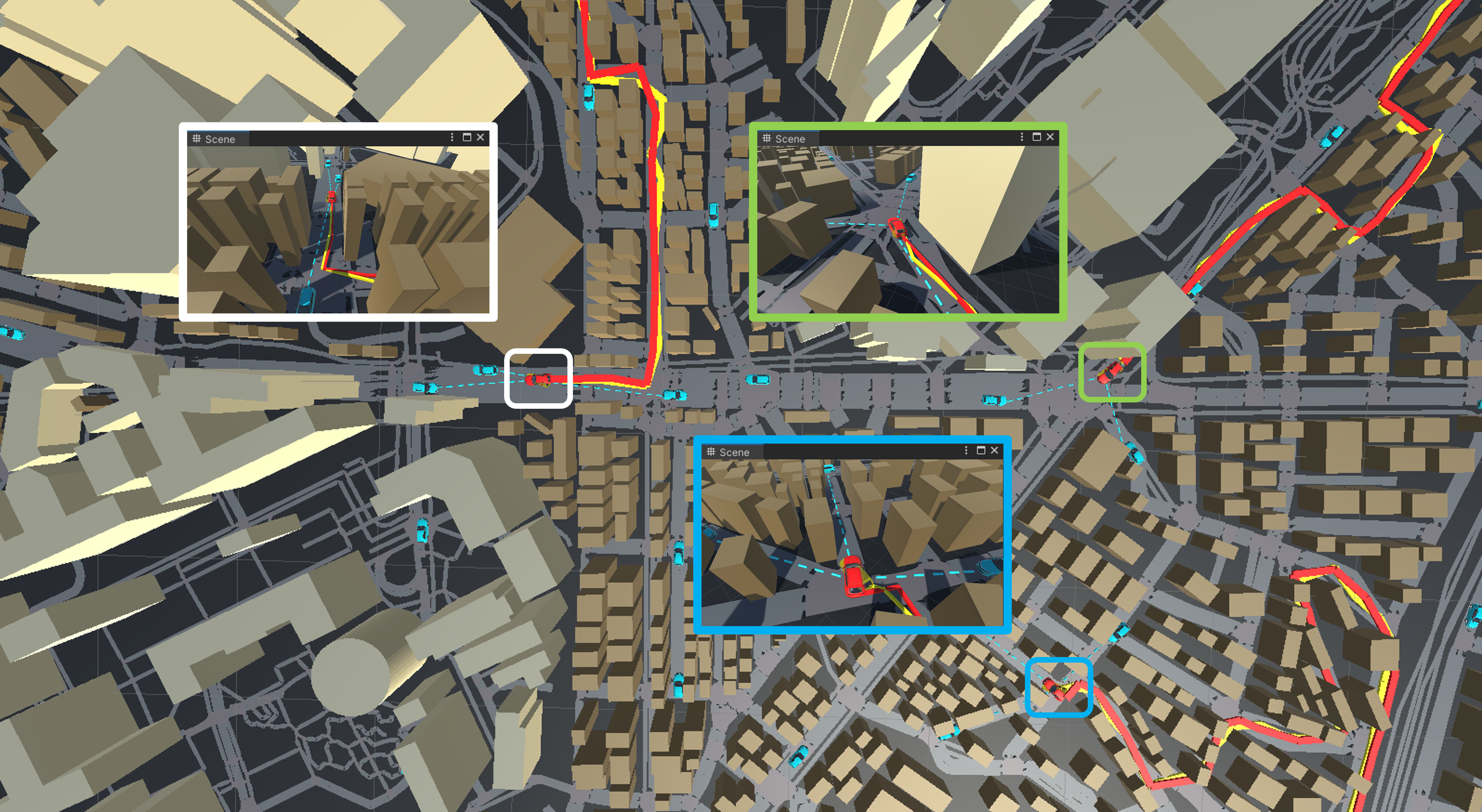}
\caption{Virtual testbed and realistic vehicular simulation in urban environment.}
\label{fig:fig4}
\end{figure*}

Fig. \ref{fig:fig4} illustrates the GUI-based testbed implementing urban environmental, vehicular mobility patterns, and developed MLCL engines. In the birds-eye view environment, instantaneous vehicular positions are represented with blue objects at their ground-truth location and orientation. Vehicles move alongside roads according to mobility trajectories in sync with the global time reference. Trajectory estimates for three selected vehicles, highlighted in red on the map, capture history of positions in red curves. Instantaneous communication links with surrounding vehicles are tracked with blue lines. For monitoring of mobility and interaction among vehicular neighborhood, the highlighted vehicles are also tracked in real-time third-person views, displayed next to vehicles upon the positioning update. The instantaneous position estimate of each vehicle, illustrated together as a twin object of yellow transparent shape, shows how the localization results infer close to reality.

\subsection{MLCL training configurations}
The architectures of DNN units are configured in the following way. Two-layer FNNs are employed for MRNN, MTNN, and LENN units with rectified linear unit activations at hidden layers. The dimension of hidden layers is set to $1000$, while the output dimension of MRNN and MTNN units is $100$. Furthermore, state updates for SUNN unit are implemented through single-layer gated recurrent unit (GRU), with state vector $S_{A}^{t}$ of length $100$ fed as its hidden state. Neural parameters are trained with Adam optimizer of learning rate $0.0005$ with $200$ mini-batch samples. Each mini-batch sample contains true instantaneous positions of $20$-second intervals for $10$ vehicles chosen from 354 vehicles in the training dataset. Internal and external measurements are prepared with noisy observation of mini-batch samples. Internal measurement $I_{A}^{t}$ contains a raw estimate on instantaneous position $X_{A}^{t}$ collected through on-board GNSS sensors. External measurements obtained in interconnected vehicle pairs include noisy observations of inter-vehicle distance and orientation \cite{SWK:21}. 
The standard error of internal measurement is $10$m, which is common in urban traffic environments where surrounding obstacles prevent reliable GNSS measurements \cite{Lin}. The average errors of inter-vehicle distance and inter-vehicle orientation are $3$m and $1^{\circ}$, respectively. Inter-vehicle interaction through V2X communication link is taken only among nearby vehicles. Thus, vehicles can interact with neighbors within $500$m and $1000$m ranges for measurement and communication domains, respectively. A random communication failure disconnects wireless inter-vehicle links with probability $0.1$, which reflects the V2X connection requirement over dense urban environments \cite{VNC18}.

\subsection{Performance evaluation}

The MLCL approach is tested for evaluating the training performance with test samples collected from the virtual testbed. The results are compared with the following five benchmark schemes:
\begin{itemize}
\item \textit{Maximum likelihood \cite{SWK:21}:} Positioning estimation proceeds in non-casual manners with all measurements collected during $20$-second intervals. The centralized optimization minimizes the joint likelihood function of internal and external measurements for positioning solutions.
\item \textit{GCN \cite{WYan:21}:} GCN models contain two subsequent graph-convolution layers capturing inter-vehicle interactions of measurement and communication domains, respectively. 
Positioning estimates of individual vehicles are processed through built-in LENN units.

\item \textit{EKF:} Centralized EKF algorithm jointly tracks all vehicular positions. Thus, perfect knowledge of all statistical information about external measurements and vehicle mobility is necessary. 
\item \textit{No cooperation (NC):} The MLCL models are trained without inter-vehicle interactions of communication domain.
\item \textit{Naive:} Positioning solutions are determined only with the internal measurement, i.e., the estimation with raw data from on-board sensors. 
\end{itemize}
The maximum likelihood technique presumes to share all internal/external measurement to obtain positioning estimates of all vehicles during 20-second intervals. The obtained result places the lower bound for localization error performance. The comparison with collaborative results can assess the impact of inter-vehicle interactions in localization tasks. GCNs have no such components as SUNN units and cannot process time domain operations. Thus, the proposed time domain design is compared with GCN techniques. All ML-based benchmarks are configured with ML architectures having similar levels of the DNN model complexity with MLCL framework.

\begin{figure}
\centering
    \subfigure[ MAE performance with respect to time.]{
        \includegraphics[width=.6\linewidth]{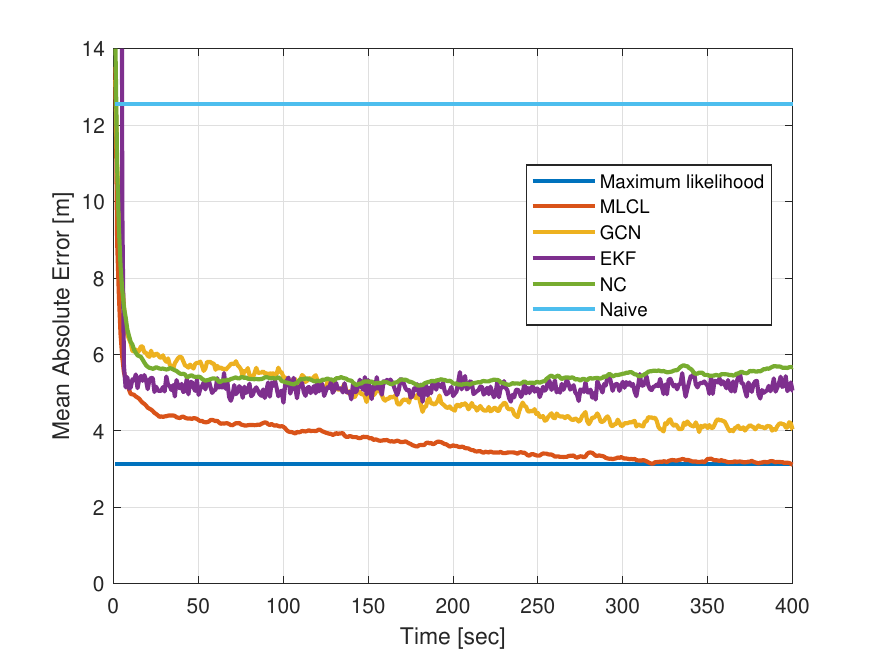}
    }
    \subfigure[ MAE performance with respect to the number of vehicles.]{
        \includegraphics[width=.47\linewidth]{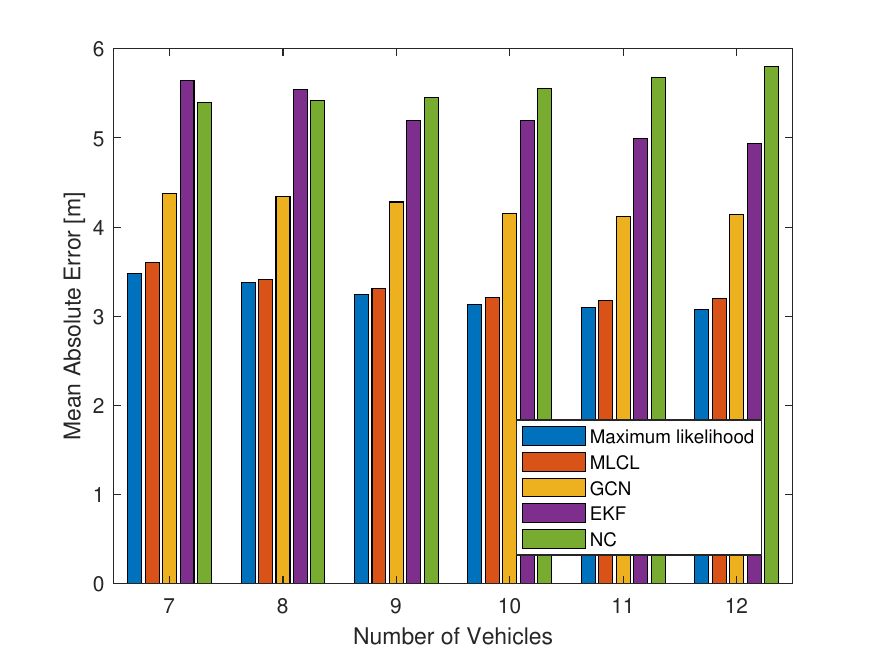}
    }
    \subfigure[MAE performance with respect to communication range.]{
        \includegraphics[width=.47\linewidth]{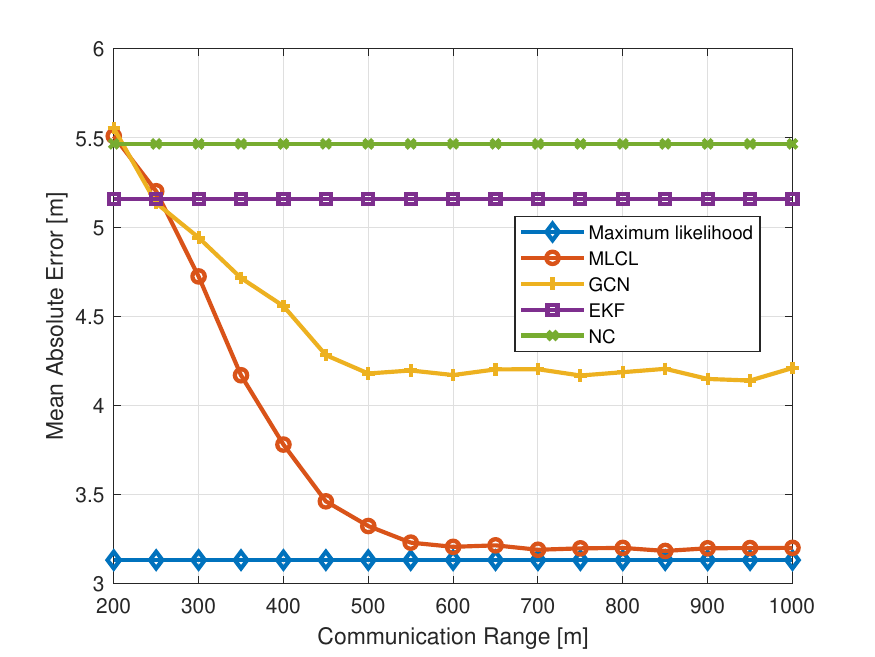}
    }
    \caption{Performance comparison of various localization schemes.}
    \label{fig:fig5}
\end{figure}

Fig. \ref{fig:fig5} compares various aspects of the localization performance with the benchmarks. The convergence behaviors of training curves are illustrated with respect to driving times in Fig. 5(a). The MLCL technique shows a rapid convergent pattern to the lower bound. It is noted that the training computation of the DNN models with 20-second mini-batch samples achieves the error bound with test samples for 400 seconds. This assures temporal scalability so that a short-duration temporal dataset suffices to train the MLCL over an extended period. Cooperative techniques obtain reduced estimation errors as compared to non-cooperative counterparts. This indicates that inter-vehicle interaction is the enabling feature of the decentralized localization. The superior performance over GCN models proves state-updating ML architecture viable in realizing intra-vehicle interactions.

The impact of collaborating vehicle population is plotted in Fig. 5(b). The MAE performance is evaluated with the average of previous results for the last 100 seconds. To validate the generalization ability, all ML-based models are trained with 10 interacting vehicles and subsequently tested with different numbers of vehicles. The MLCL framework shows consistent performance superiority over all ranges of neighborhood populations. Noticeably, the localization performance gradually improves even with 11 and 12 neighbors, exceeding 10 vehicles. Thus, the proposed strategy works well in an extended configuration that has not been considered in the training computation. Furthermore, it exhibits competitive performance with fully centralized maximum likelihood scheme, thereby proving the generalization ability. By contrast, the MAE performance of NC method worsens since it does not benefit from communication messages. This indicates that communication messages become valuable for increasing number of interacting neighbors.

The coverage of inter-vehicle communication varies with carrier frequency and antenna specifications. For the impact of parameter configurations, Fig. 5(c) shows the MAE performance with respect to the communication range when the measurement range is fixed to $500$m. The increasing communication range yields monotonic improvement of the localization performance both with MLCL and GCN. In particular, initial MAE performance of the MLCL framework as large as the non-cooperation result reveals gradual improvement to reach the best lower-bound performance. Also, the communication range of $600$m suffices to achieve the best performance in urban environment. By contrast, GCN suffers from the error floor, failing to obtain the lower-bound error performance as the communication environment scales up. 
These results validate the MLCL approach as a successful inter-vehicle collaborative framework for accurate vehicular positioning.

\section{Conclusion and technical challenges}
This article consolidates the feasibility of ML-based collaborative localization under urban traffic environments. 
This ML-aided autonomous framework slices vehicular networks into three different interaction domains, each processed with distinct DNN units. Individual units are responsible for different types of vehicular interaction dedicated to those domains, and their coordination establishes sophisticated localization solutions. The performance of MLCL framework is validated over urban map based testbed. 
To elaborate practical deployment of the proposed framework, the following issues need further investigation:

\begin{itemize}
    \item Collective driving control: Integrated design of autonomous localization and control further enhances collective safety of vehicular groups. To this end, digital-twin platform for autonomous driving provides solid data collection and reliable simulation-based test for ML-assisted operations.
    \item Measurement reliability: Knowledge of non-LoS statistics improves measurement reliability. Therefore, additional neural networks can obtain informative observations from pilot signaling exchanged over measurement domain.
    \item Security against adversary: ML-assisted cooperative techniques presume benign vehicles. DNN units implementing collaborative localization are susceptible to adversarial attacks by stranger vehicles. This poses secure autonomous controls that avoid adversarial inputs to manage sustainable collaborative localization.  \item Complexity reduction: Timely localization using ML models requires reducing the DNN model complexity, with high localization accuracy preserved. Knowledge distillation techniques can be viable to reduce the model complexity.
\end{itemize}

Supported by future research efforts about collaborative localization, MLCL framework can pave a way towards urban roadway deployment.

\bibliography{arxiv}
\bibliographystyle{ieeetr}

\end{document}